\documentclass[12pt]{iopart}
\usepackage{graphicx}
\usepackage{amssymb}
\usepackage{bm}
\usepackage{slashbox}

\begin{document}
\title{Impurity cyclotron resonance of anomalous  Dirac electrons in graphene}
\author{S. C. Kim$^{1}$,  S. -R. Eric Yang$^{1,\dag}$, and   A. H. MacDonald$^{2}$}
\address{$^{1}$ Physics Department, Korea  University, Seoul 136-713, Korea\\
$^{2}$ Physics Department, University of Texas, Austin \\
}
\ead{$^{\dag}$corresponding author, eyang812@gmail.com}

\begin{abstract}
We have investigated a new feature of impurity cyclotron resonances
common to various localized potentials of graphene.  A localized
potential can interact with a magnetic field in an unexpected  way
in graphene. It can lead to formation of  anomalous boundstates that
have a sharp peak with  a width $R$ in  the probability density
inside the potential and a broad peak of size magnetic length $\ell$
outside the potential.  We investigate optical matrix elements of
anomalous states, and find that they are unusually small and depend
sensitively on magnetic field. The effect of many-body interactions
on their optical conductivity is investigated using a
self-consistent time-dependent Hartree-Fock approach (TDHFA). For a
completely filled Landau level we find that an  excited
electron-hole pair, originating from the optical transition between
two anomalous impurity states, is nearly uncorrelated with other
electron-hole pairs, although it displays a substantial exchange
self-energy effects. This  absence of correlation is a consequence
of  a small vertex correction in comparison to the difference
between renormalized transition energies computed within the one electron-hole pair approximation.  However, an  excited
electron-hole pair originating from the optical transition between a
normal and an anomalous impurity states can be substantially
correlated with other electron-hole states with a significant
optical strength.
\end{abstract}
\maketitle

\section{Introduction}

A Dirac electron\cite{Mac0,Ben,Castro} moving in a rotationally
invariant, localized, and smoothly varying potential $V(r)$  is
described by the Hamiltonian
\begin{eqnarray}
H=v_F\vec{\sigma}\cdot(\vec{p}+\frac{e}{c}\vec{A})+V(r),
\end{eqnarray}
where  $\vec{\sigma}=(\sigma_x,\sigma_y)$ and $\sigma_{z}$ are Pauli
spin matrices ($\vec{p}$ is two-dimensional momentum). The shape of $V(r)$ can be parabolic,
Coulomb, and Gaussian.   Half-integer angular momentum $J$ is a good
quantum number and wavefunctions of eigenstates have the form
\begin{equation}
\Psi^{J}(r,\theta)=\left(
\begin{array}{c}
\chi_{A}(r)e^{i(J-1/2)\theta}\\
\chi_{B}(r)e^{i(J+1/2)\theta}
\end{array}
\right).\label{wavef}
\end{equation}
It consists of  A sublattice and B sublattice  radial wavefunctions
$\chi_A(r)$ and $\chi_B(r)$ with channel angular momenta $J-1/2$ and
$J+1/2$, respectively. The half-integer angular momentum quantum
numbers have values $J=\pm1/2,\pm3/2,\cdots$.  The Hamiltonian has
several unusual features   not present in the case of massful
electrons. A simple scaling analysis\cite{Jac} suggests that a
localized potential $V(r)$ can act as a strong perturbation, and
that it can be even more singular in graphene than in ordinary
two-dimensional systems of massful electrons: the kinetic term of
Dirac Hamiltonian scales as $1/r$ while the potential term scales as
$1/r$ and $1/r^2$ for  Coulomb and Gaussian short-range
potentials\cite{com0}, respectively.  The other unusual feature is
the presence of quasibound states with complex
energies\cite{Mat,gia}.

\begin{figure}[!hbpt]
\begin{center}
\includegraphics[width=0.3\textwidth]{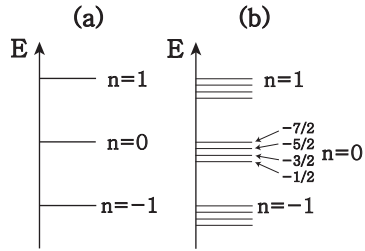}
\caption{ (a) Schematic drawing of degenerate LL energies of
graphene.   LL energy is $nE_M $, where $n=...-1,0,1,...$ is LL
index and $E_M=\hbar v_F/\ell=26(B\textrm{[T]})^{1/2}\textrm{[meV]}$
is the characteristic energy scale of graphene LLs. (b) When an
attractive localized potential is present LL energies split into
discrete energies. Fractions indicate values of angular momentum $J$.} \label{splitLL}
\end{center}
\end{figure}

\begin{figure}[!hbpt]
\begin{center}
\includegraphics[width=0.45\textwidth]{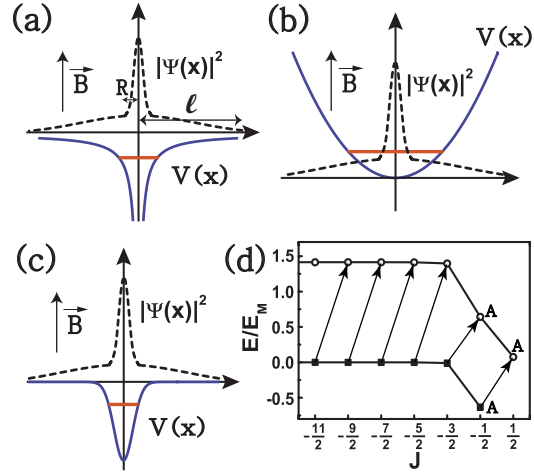}
\caption{Schematic display of probability density of anomalous
states (small peak has width $R$ and broad peak $\ell$). (a)
regularized Coulomb potential, (b) parabolic potential, and (c)
Gaussian potential. Horizontal lines indicate the energy levels. (d)
Energy splitting of $n=0$ (chiral) and $1$ (nonchiral) LLs is shown.
Filled squares represent occupied $n=0$ LL states and open circles
represent empty $n=1$ LL states. Symbol A stands for an anomalous
state.  Arrows indicate optical transitions.
  } \label{anoma}
\end{center}
\end{figure}

These effects show up differently  in  a magnetic field $\vec{B}$
applied perpendicular to the two-dimensional plane  (the vector
potential $\vec A$ is given in a symmetric gauge).  In the absence
of $V(r)$ eigenenergies form degenerate Landau level (LL) energies
while they split into discrete energies when $V(r)$ is present, see
Fig.\ref{splitLL}. They can form true boundstates with real
energies\cite{gia,Rec,Sch} in contrast to the case of no magnetic
field.  Moreover, in addition to the magnetic length,
$\ell=25.66(B\textrm{[T]})^{-1/2}\textrm{[nm]}$, a new length scale
$R$ is introduced in the wavefunction: boundstates with a s-channel angular momentum
component  can become {\it anomalous} and develop a sharp peak of a
width $R$ inside the potential and a broad peak of size magnetic
length $\ell$ outside the potential\cite{Park}. Although the effect
of the potential is strong it is partly mitigated by Klein
tunneling, and there is a competition between the two length scales
$R$ and $\ell$: the peak is strong in the regime $R/\ell<1$, but
small  in the regime $R/\ell>1$ (in the limit $R/\ell\rightarrow 0$
it diverges).  These states are present in various potentials:
regularized Coulomb\cite{Kim0,Ho}, parabolic\cite{gia,Kim2}, and
finite-range potentials\cite{Rec,Park}, see Fig.\ref{anoma}(a), (b),
and (c).

However, experimentally it is unclear how to probe these anomalous
states.  We propose in this paper that they can lead to a new
feature in  impurity cyclotron resonances between  boundstates of
LLs\cite{Jar,Gold}, shown in Fig.\ref{anoma}(d). Optical properties
in a magnetic field have several new features not present in   a
quantum dot in zero magnetic field\cite{zero}: the bulk  $n=0$ LL
states are all chiral (one-component) while   the states of other LLs
are nonchiral (two-component). As a consequence, the $n=0$ LL has
only one anomalous state while the $n=1$ LL has two, see
Eq.(\ref{wavef}). The optical transitions originating from these
states depicted in Fig.\ref{anoma}(d). Another unique feature of
optical transitions involving anomalous boundstates is that their
optical matrix elements  are rather {\it small} since their
wavefunctions are peaked at $r=0$ (this is explicitly shown in
Sec.2). Their optical conductivity is thus smaller than those of the
usual boundstates. In addition, since the optical strength depends
{\it sensitively} on magnetic field, we suggest that  a magnetic
field can serve as an experimental control parameter for detecting
the new magnetospectroscopic feature. The electron-electron Coulomb
interaction\cite{many} may also affect these transitions.  The
energy scale of the electron-electron Coulomb interaction is
significant in graphene at all values of magnetic fields
$B$\cite{com1}, and each optically excited  state is expected to be
a correlated  many-body state, containing a linear combination of
several electron-hole pair states.  This effect may change the value
of the optical strength. We have investigated this issue for a
completely filled LL by computing many-body correlated states within
TDHFA\cite{TDHF3,TDHF2,TDHF1,Bychkov,Pfa} and have calculated the
optical conductivity. In contrast to the naive expectation, we find
that an  excited electron-hole pair originating from the optical
transition between  two anomalous impurity states (see
Fig.\ref{anoma}) is nearly uncorrelated with other electron-hole
states, despite displaying substantial exchange self-energy effects.
This  absence of correlation is a consequence of  a small vertex
correction in comparison to the difference between renormalized
transition energies computed within the one electron-hole pair approximation.  Many-body interactions do not  enhance the
strength of the optical conductivity of this transition.  However,
an  excited electron-hole pair originating from the optical
transition between a normal and an  anomalous impurity states (see
Fig.\ref{anoma}) can be substantially correlated with other
electron-hole pairs with a significant optical strength.

This paper is organized as follows. In Sec.2  the optical matrix
element of anomalous states is shown to be small using  an idealized
impurity model.  The many-body version of the Kubo formula of
optical conductivity is given in Sec.3. How the self-energy
correction of a singly occupied LL state affects the optical
conductivity is  evaluated in Sec.4.  In Sec.5 we investigate how
important  vertex corrections are  for a completely filled LL. The
final section 6 includes a summary and discussion.

\section{Model potential: optical matrix elements of anomalous states }

In order to compute the optical conductivity of anomalous states
accurately we first solve exactly the single impurity problem, and
apply the TDHFA using these solutions. A similar method was used for
massful electrons confined in a quantum dot\cite{Pfa}. We study
optical properties of anomalous states using a simple model
potential.  We choose a cylindrical impurity potential\cite{Rec}
since its eigenstates and eigenvalues can be solved {\it exactly} in
the presence of a magnetic field. This potential captures essential
features of anomalous states of parabolic, Coulomb, and Gaussian
potentials. The potential has the shape
\begin{eqnarray}
V_I(\vec{r})=\left\{\begin{array}{c}V_I \ \ \ \ r<R\\
0 \ \ \ \ \ r>R\end{array}\right. ,\label{eqV}
\end{eqnarray}
where $V_I$ is the strength of the potential and $R$ is the radius
of the cylinder. According to Eq.(\ref{wavef}) the impurity state
$|n,J\rangle$ of $n$th LL with {\it conserved} angular momentum
quantum number $J$ has the following form of wavefunction
$\Psi_n^J(\vec{r})$
\begin{eqnarray}
\langle \vec{r}|n,J\rangle=\Psi_n^J(r,\theta)=\left(\begin{array}{c}\chi^{l_A}_{n,A}(r)e^{i(J-1/2)\theta}\\
\chi^{l_B}_{n,B}(r)e^{i(J+1/2)\theta}\end{array}\right),\label{eigenstates}
\end{eqnarray}
where the A-component has {\it orbital} angular momentum $l_A=J-1/2$
and B-component $l_B=J+1/2$. The radial wavefunctions are given in
\cite{Rec,Yosi}
\begin{eqnarray}
&&\chi^{l_{\sigma}}_{n,\sigma}(r)
\sim e^{\frac{-br^{2}}{2}}r^{n_{\sigma}}\left\{\begin{array}{c}\alpha_{\sigma}U(q_{\sigma},1+n_{\sigma},br^{2}), \ \ \ r>R\\
\beta_{\sigma}M(q_{\sigma},1+n_{\sigma},br^{2}), \ \ \ r<R\end{array}\right. ,\nonumber\\
\end{eqnarray}
where the sublattice index $\sigma=A,B$ and $U(x,y,z)$ and
$M(x,y,z)$ are confluent hypergeometric functions. The {\it impurity
eigenenergy} is denoted by $\tilde{\epsilon}_{n,J}$, and is plotted
as a function of $J$  for LL indices $n=0$  and $1$ in
Fig.\ref{anoma}(d).

\begin{figure}[!hbpt]
\begin{center}
\includegraphics[width=0.45\textwidth]{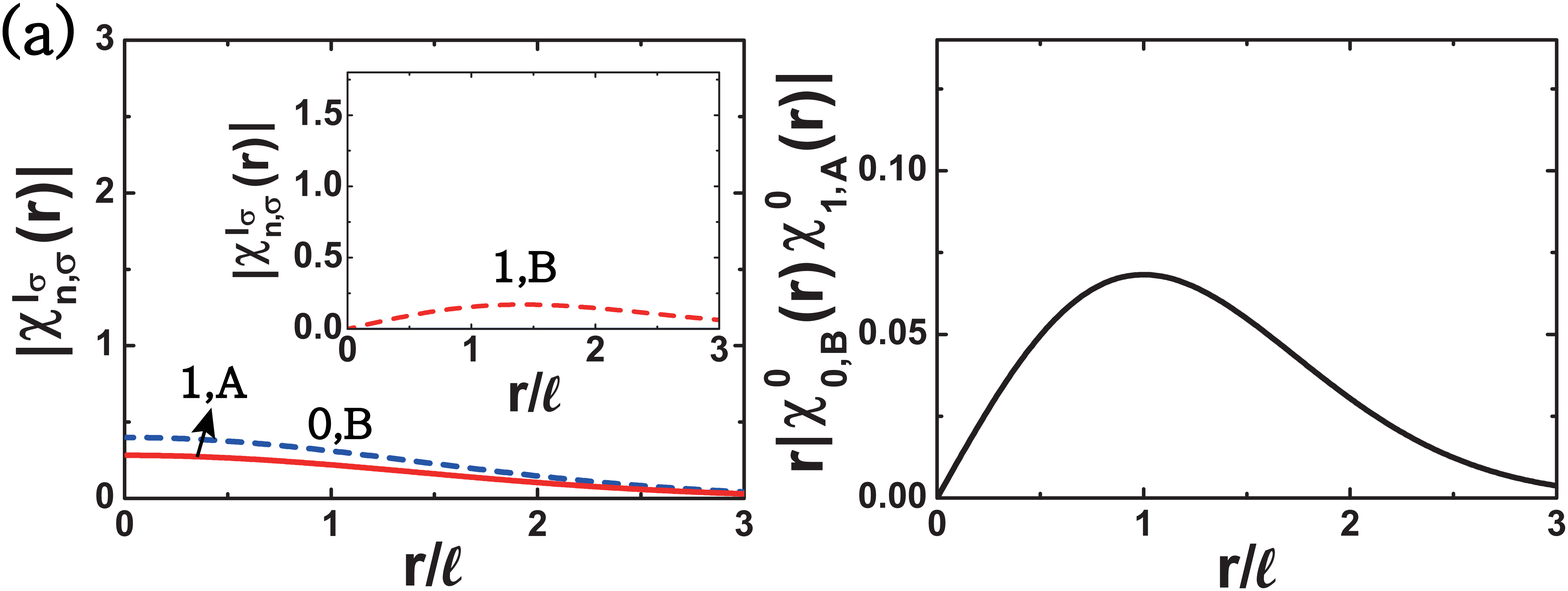}
\includegraphics[width=0.45\textwidth]{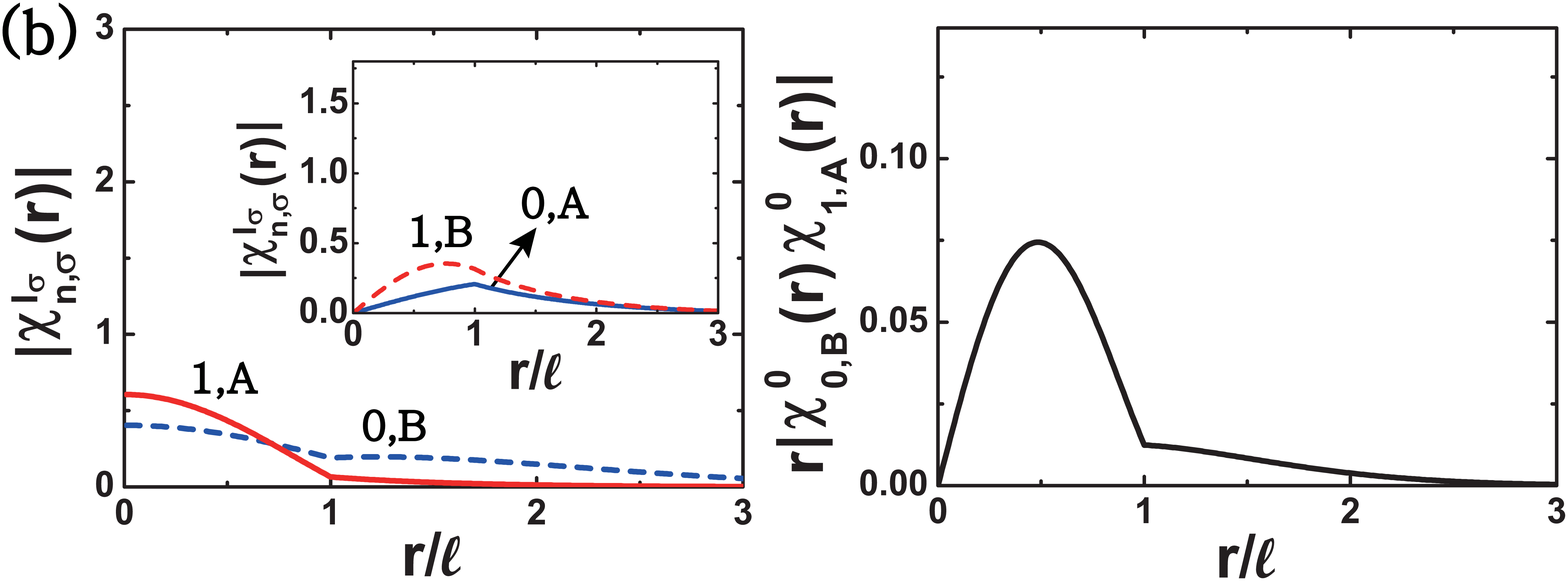}
\includegraphics[width=0.45\textwidth]{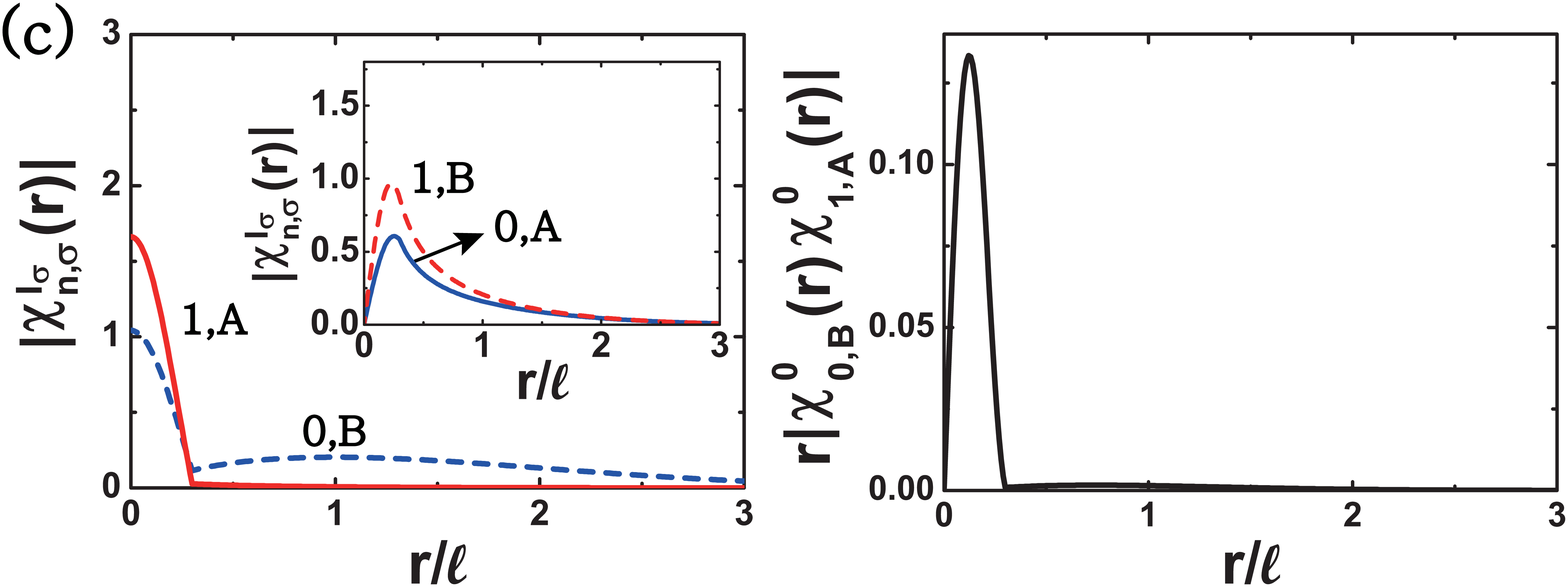}
\includegraphics[width=0.45\textwidth]{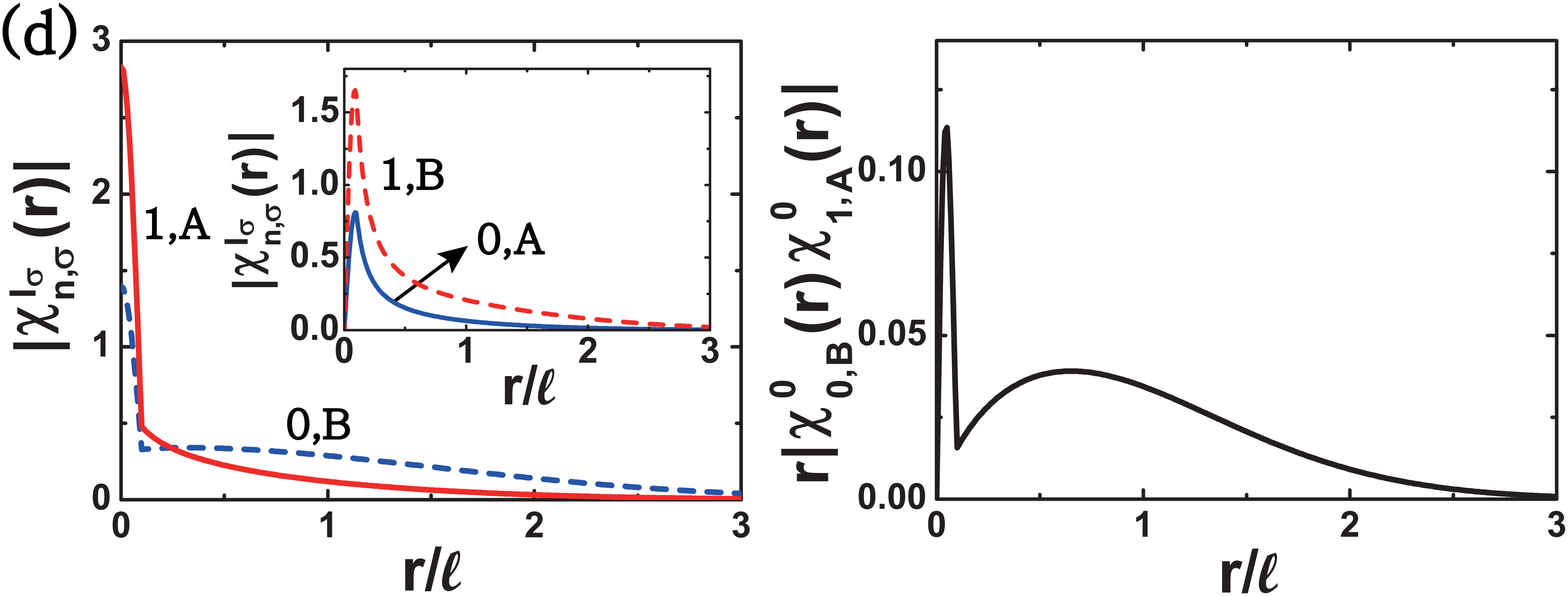}
\caption{Wavefunction properties of impurity states $|0,-1/2
\rangle$ and $|1,1/2 \rangle$. Left column: s-wave radial functions
$|\chi_{0,B}^{0}(r)|$ (dashed line) and $|\chi_{1,A}^{0}(r)|$ (solid
line).  Inset: p-wave radial functions $|\chi_{0,A}^{-1}(r)|$ (solid
line) and $|\chi_{1,B}^{1}(r)|$ (dashed line). From (a) to (d) the
values of parameters are $(\frac{V_I}{E_M},\frac{R}{\ell})=(0,0),
(-2,1), (-7.9,0.3), (-20,0.1)$. Right column: integrand
$r\chi_{0,B}^{0}(r)\chi_{1,A}^{0}(r)$ is displayed. }
\label{prob_den}
\end{center}
\end{figure}

Let us investigate which impurity states $\Psi_n^J(\vec{r})$ can be
anomalous, i.e., which of these states can have s-wave radial
functions. It is possible only for states with $J=\pm 1/2$, see
Eq.(\ref{wavef}). Using confluent hypergeometric functions and
Eq.(\ref{eigenstates}) it can be shown that, for $n=0$ LL, only the
state $\Psi_0^{-1/2}(\vec{r})$ has value at the origin
$\Psi_0^{-1/2}(0)\neq0$ since its the B-component radial
wavefunction $\chi^{0}_{0,B}(r)$ is s-wave. While, for $n=1$ LL,
both $\Psi_1^{-1/2}(\vec{r})$ and $\Psi_1^{1/2}(\vec{r})$ can be
anomalous since their B- and A-components of radial wavefunctions
are s-wave, respectively.  These  states are labeled by $A$ in
Fig.\ref{anoma}(d).  They actually become anomalous when the
condition $R/\ell<1$ is satisfied, see Fig.\ref{prob_den}.

Photons are assumed to be polarized along x-axis, and the optical
matrix elements can be computed using  the current operator
$\vec{j}=v_F\vec{\sigma}$\cite{Ben}. They satisfy the selection
rules with the change of LL index $\Delta n=1$ and change of angular
momentum $\Delta J=1$\cite{optic}, and the relevant single-particle
optical transitions are of the type $|0,J\rangle\rightarrow
|1,J+1\rangle$.  The optical matrix element between the relevant
anomalous states can be expressed in terms of the corresponding
radial wavefunctions (see Eq.(\ref{eigenstates}))
\begin{eqnarray}
\langle 1,1/2|\sigma_{x}|0,-1/2\rangle &=&
\langle e^{i\theta}\chi^{1}_{1,B}| e^{-i\theta}\chi^{-1}_{0,A}\rangle+\langle \chi^{0}_{1,A}| \chi^{0}_{0,B}\rangle\nonumber\\
&=&\langle \chi^{0}_{1,A}| \chi^{0}_{0,B}\rangle.
\end{eqnarray}
In order to understand why this optical matrix element can be small
the radial wavefunctions $\chi^{0}_{1,A}(r)$ and $\chi^{0}_{0,B}(r)$
in $\langle \chi^{0}_{1,A}| \chi^{0}_{0,B}\rangle$ are plotted,
together with the integrand of $\int dr r
\chi^{0}_{1,A}(r)^{*}\chi^{0}_{0,B}(r) $,  in Fig.\ref{prob_den}.
We see from the shape integrand, Figs.\ref{prob_den}(b), (c), and
(d), that the resulting integral is {\it smaller} than the
corresponding integral in the absence of the localized potential
plotted in Fig.\ref{prob_den}(a). The actual values of transition
matrix elements $|\langle 1,J|\sigma_{x}|0,J'\rangle|^{2}$ are given
in Table \ref{table} for varies values of
$\Big(\frac{V_{I}}{E_{M}},\frac{R}{\ell}\Big)$. The dependence of
the  transition matrix elements on the parameters
$\Big(\frac{V_{I}}{E_{M}},\frac{R}{\ell}\Big)$ is non-trivial.  Note
that they depend {\it significantly} on the ratio $R/\ell$, i.e., on
magnetic field.

\begin{table}[!hbpt]
\caption{Transition matrix elements $|\langle
1,J|\sigma_{x}|0,J'\rangle|^{2}$} \centering
\begin{tabular}{|c|c|c|c|c|c|}
\hline \textrm{\tiny{\backslashbox{$(V_I/E_{M},R/\ell)$}{$(J,J')$}}}
& \textrm{\tiny{$(\frac{1}{2},-\frac{1}{2})$}} &
\textrm{\tiny{$(-\frac{1}{2},-\frac{3}{2})$}} &
\textrm{\tiny{$(-\frac{3}{2},-\frac{5}{2})$}}
& \textrm{\tiny{$(-\frac{5}{2},-\frac{7}{2})$}} & \textrm{\tiny{$(-\frac{7}{2},-\frac{9}{2})$}}  \\
\hline $(-2,1)$ & 0.137 & 0.263 & 0.377 & 0.487 & 0.5 \\
\hline $(-7.9,0.3)$ & 0.025 & 0.146 & 0.5 & 0.5 & 0.5 \\
\hline $(-20,0.1)$ & 0.157 & 0.341 & 0.5 & 0.5 & 0.5 \\
\hline
\end{tabular}\label{table}
\end{table}

\section{Optical conductivity}

We compute the many-body optical conductivity in the presence of a
single impurity (when more impurities are present in the dilute
limit the total optical conductivity is given by the sum of the
optical conductivity of  each impurity\cite{Yang}). We consider
excitations from a singly occupied LL or completely filled LL
(partially filled LLs cannot be described adequately in TDHFA since
screening becomes important). The groundstate is denoted by
$|\psi_F\rangle$, and it represents either a singly occupied LL or
completely filled LL. In this case the optical conductivity consists
of a series of discrete peaks.

Generally a many-body excited state can be written as a linear
combination of single-electron excited states
\begin{eqnarray}
|\Psi\rangle=\sum_J  C_J |\psi_{J}\rangle, \label{linear-comb}
\end{eqnarray}
where single-electron excited states with $\Delta J=1$ are
\begin{eqnarray}
|\psi_{J}\rangle =a^{\dag}_{n',J+1}a_{n,J}|\psi_F\rangle.
\label{basis}
\end{eqnarray}
Here the operator  $a^{\dag}_{n,  J}$ creates an electron in the
localized state of $n$th LL with angular momentum $J$.  The optical
conductivity consists of a series of discrete peaks at the
renormalized excitation energies $E_{ex}$
\begin{eqnarray}
\sigma(E)=\sum_{E_{ex}}s(E)\delta(E-E_{ex}),
\end{eqnarray}
where   $s(E)$ is the optical strength. When this strength is
divided by a constant $c=1/(\sqrt{2}E_M)$ it has {\it dimension of
conductivity} (if other value of $c$ is chosen the magnitude of this
scaled conductivity will be different). The scaled optical strength
is computed using the Kubo formula
\begin{eqnarray}
\tilde{s}(E_{ex})=\frac{s(E_{ex})}{1/(\sqrt{2}E_M)}=\frac{\pi}{2}
\frac{e^2}{h}  \frac{|\langle \Psi|T|\psi_F\rangle |^2}{
\tilde{E}_{ex} },\label{eq:scaled}
\end{eqnarray}
where the scaled excitation energy is $\tilde{E}_{ex}=E_{ex}/
(\sqrt{2}E_M)$. When photons are polarized along x-axis the optical
matrix element is given by $\langle \Psi|T|\psi_F\rangle$, where
\begin{eqnarray}
T=\sum_{J}\langle n',J+1|\sigma_{x}|n,J\rangle
a^{\dag}_{n',J+1}a_{n,J}+h.c.
\end{eqnarray}
The computed the optical matrix element for $n=0$ and $n'=1$ LLs can
be written in terms of expansion coefficients $C_J$ and the optical
many-body matrix elements
\begin{eqnarray}
\langle \Psi| T
|\psi_F\rangle&=&C_{-1/2}^*\langle1,1/2|\sigma_{x}|0,-1/2\rangle\nonumber\\
&+&C_{-3/2}^*\langle1,-1/2|\sigma_{x}|0,-3/2\rangle\nonumber\\
&+&C_{-5/2}^*\langle 1,-3/2|\sigma_{x}|0,-5/2\rangle\nonumber\\
&+&C_{-7/2}^*\langle 1,-5/2|\sigma_{x}|0,-7/2\rangle\nonumber\\
&+&\cdots .
\end{eqnarray}

\section{Singly occupied  Landau level}

\begin{figure}[!hbpt]
\begin{center}
\includegraphics[width=0.15\textwidth]{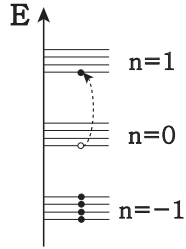}
\caption{ A single electron is in the $n=0$ LL. The LLs
$n=-1,-2,...$ are all filled.} \label{levels}
\end{center}
\end{figure}

\begin{figure}[!hbpt]
\begin{center}
\includegraphics[width=0.25\textwidth]{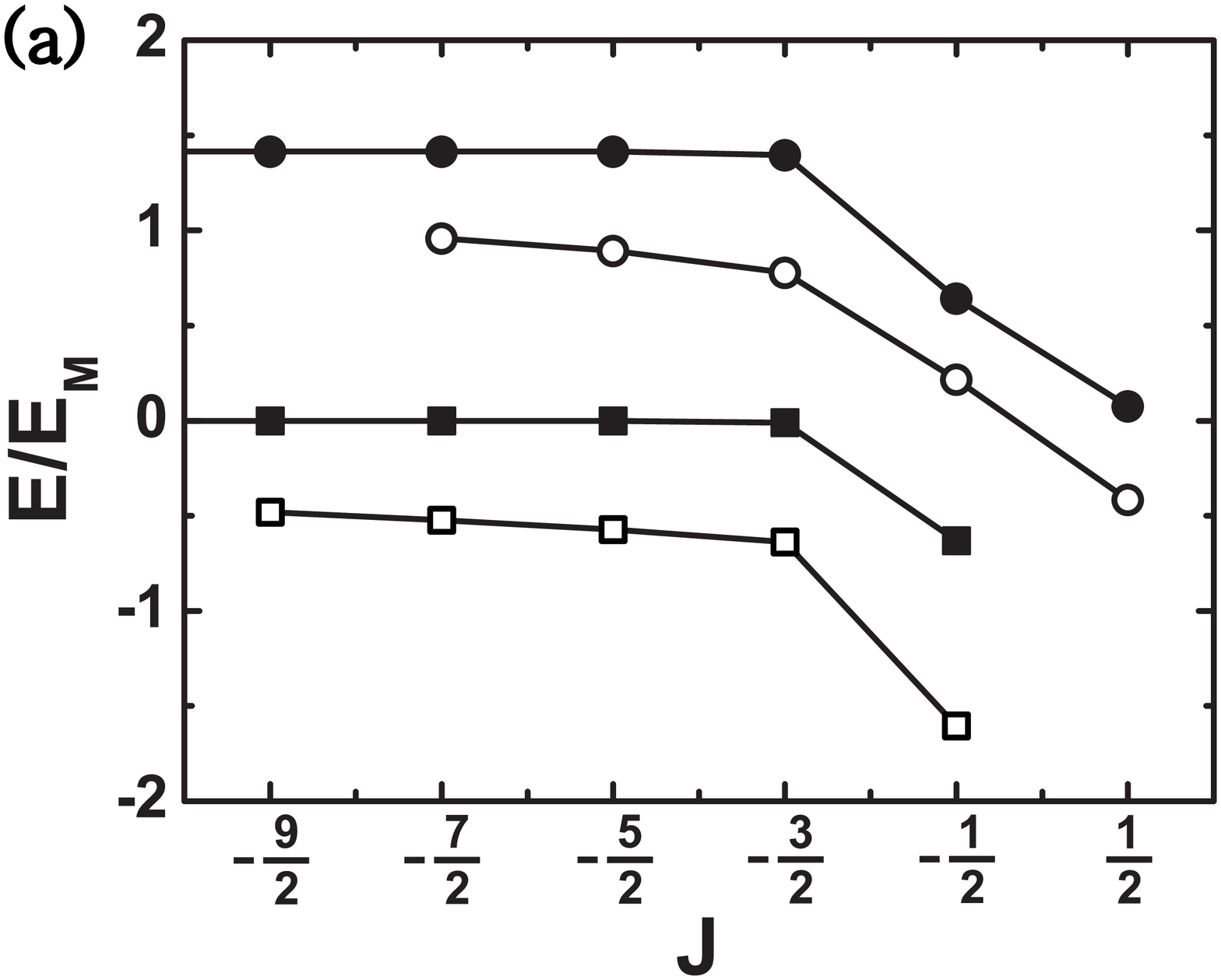}
\includegraphics[width=0.25\textwidth]{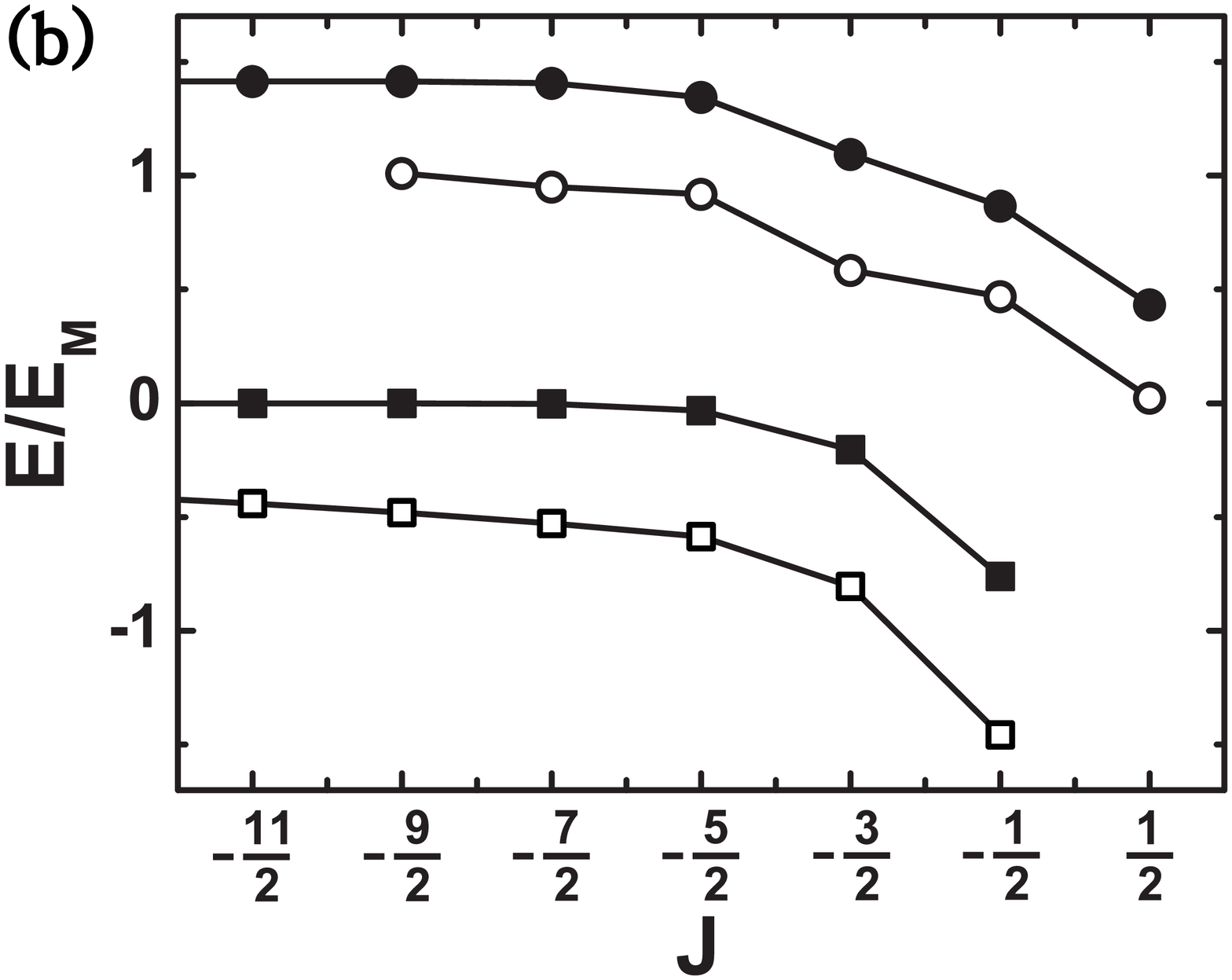}
\caption{ Bare impurity energies $\tilde{\epsilon}_{n,J}$ for $n=0$
and $1$ LL states (filled symbols). Energies corrected by
self-energy correction $E_{n,J}$ (open symbols). (a) $V_I=-7.9E_M$
and $R=0.3\ell$. (b) $V_I=-2E_M$ and $R=\ell$.} \label{selfenergy1}
\end{center}
\end{figure}

Before we investigate the effect of many-body correlations on the
optical conductivity of anomalous states we compute it without them
and include only self energy effects. This calculation is a good
approximation and is experimentally relevant when the $n=0$ LL is
occupied by only {\it one} electron. In this case only one term
survives in the linear combination given by Eq.(\ref{linear-comb}):
there is only one excited state  with $\Delta n=1$ and $\Delta J=1$,
namely $|\psi_{J}\rangle =a^{\dag}_{1,J+1}a_{0,J}|\psi_F\rangle$.
This implies that many-body correlation effects are {\it not}
present. However, self-energy corrections are {\it present}.  The
LLs with energies lower than the split $n=0$ energies, i.e., the
$n=-1,-2,...$ LLs are all occupied, see Fig.\ref{levels}. Bare
impurity  eigenenergies $\tilde{\epsilon}_{n,J}$, shown in
Fig.\ref{anoma}(d), will acquire self-energy corrections originating
from interactions with the filled electrons. The {\it renormalized
impurity  energy} is
\begin{eqnarray}
E_{n,J}=\tilde{\epsilon}_{n,J}+\Sigma_{n,J}^H+\Sigma_{n,J}^X.
\end{eqnarray}
The Hartree self-energy originates from the electronic density and
ionic potential, and has two parts
\begin{eqnarray}
\Sigma_{m,J}^{H}&=&\sum_{n',J'}f_{n',J'} \langle m,J;n',J'|V|m,J;n',J'\rangle \nonumber\\
&-& \sum_{n',l'} f_{n',l'}\langle
m,J;n',l'|V|m,J;n',l'\rangle,\label{Hartree}
\end{eqnarray}
where electron-electron interaction is
$V(r_1-r_2)=\frac{e^2}{\epsilon|r_1-r_2|}$ and the occupation
functions $f_{n,J}=1/0$ if $|n,J\rangle$ state is
occupied/unoccupied. The second term represents a correction due to
uniform ionic potential ($|n,l\rangle$ represents a LL state of
graphene in the absence of an impurity potential). The Hartree
self-energy corrections are negligibly small. The exchange
self-energy is given by
\begin{eqnarray}
\Sigma_{m,J}^{X}&=&-\sum_{n',J'} f_{n',J'} \langle
m,J;n',J'|V|n',J';m,J\rangle \label{exchange}
\end{eqnarray}
when $\mathbf{K}$ and $\mathbf{K'}$ valleys are uncoupled\cite{Lee}.
Energies corrected by these  exchange self-energy corrections are
shown in Fig.\ref{selfenergy1}.

\begin{figure}[!hbpt]
\begin{center}
\includegraphics[width=0.25\textwidth]{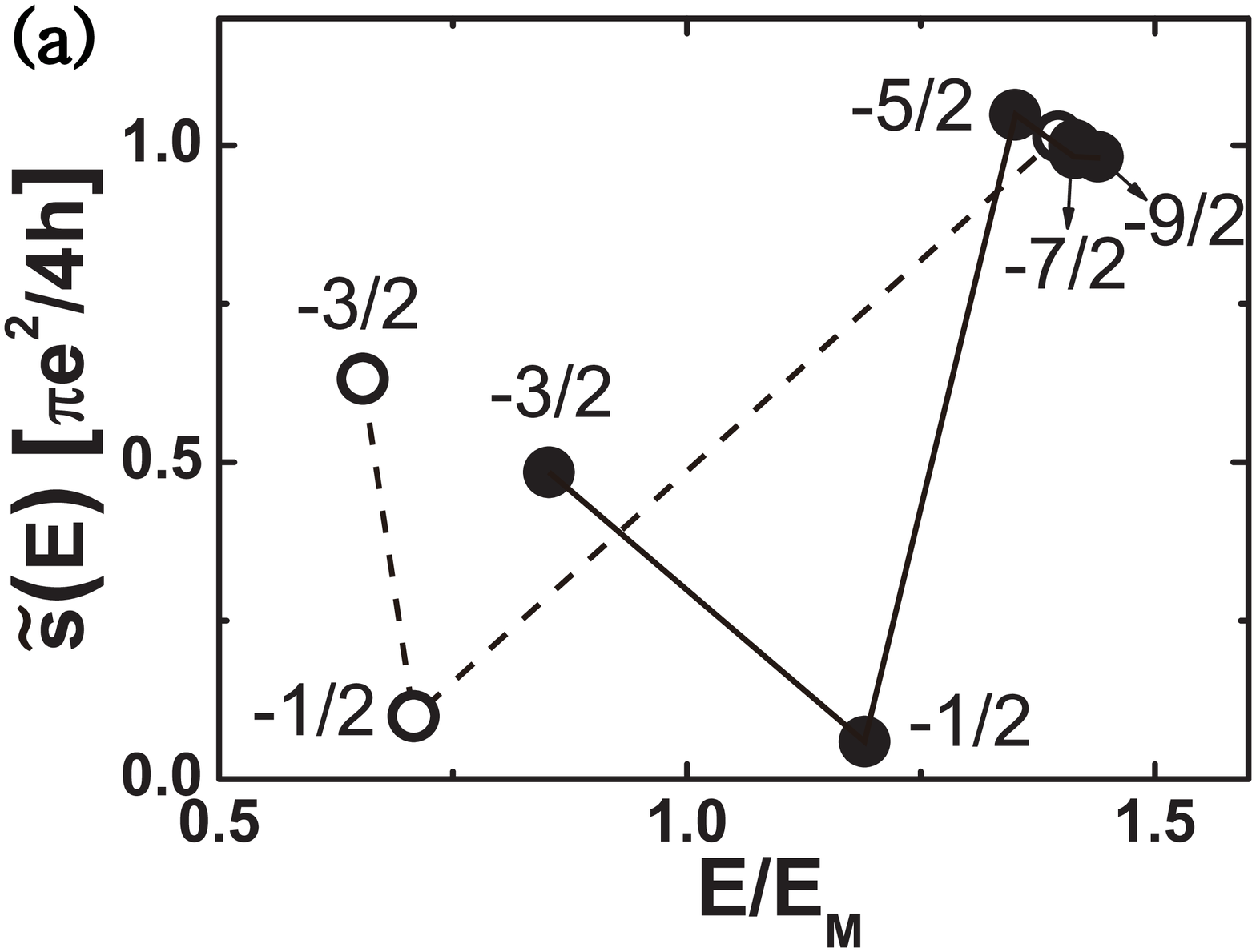}
\includegraphics[width=0.25\textwidth]{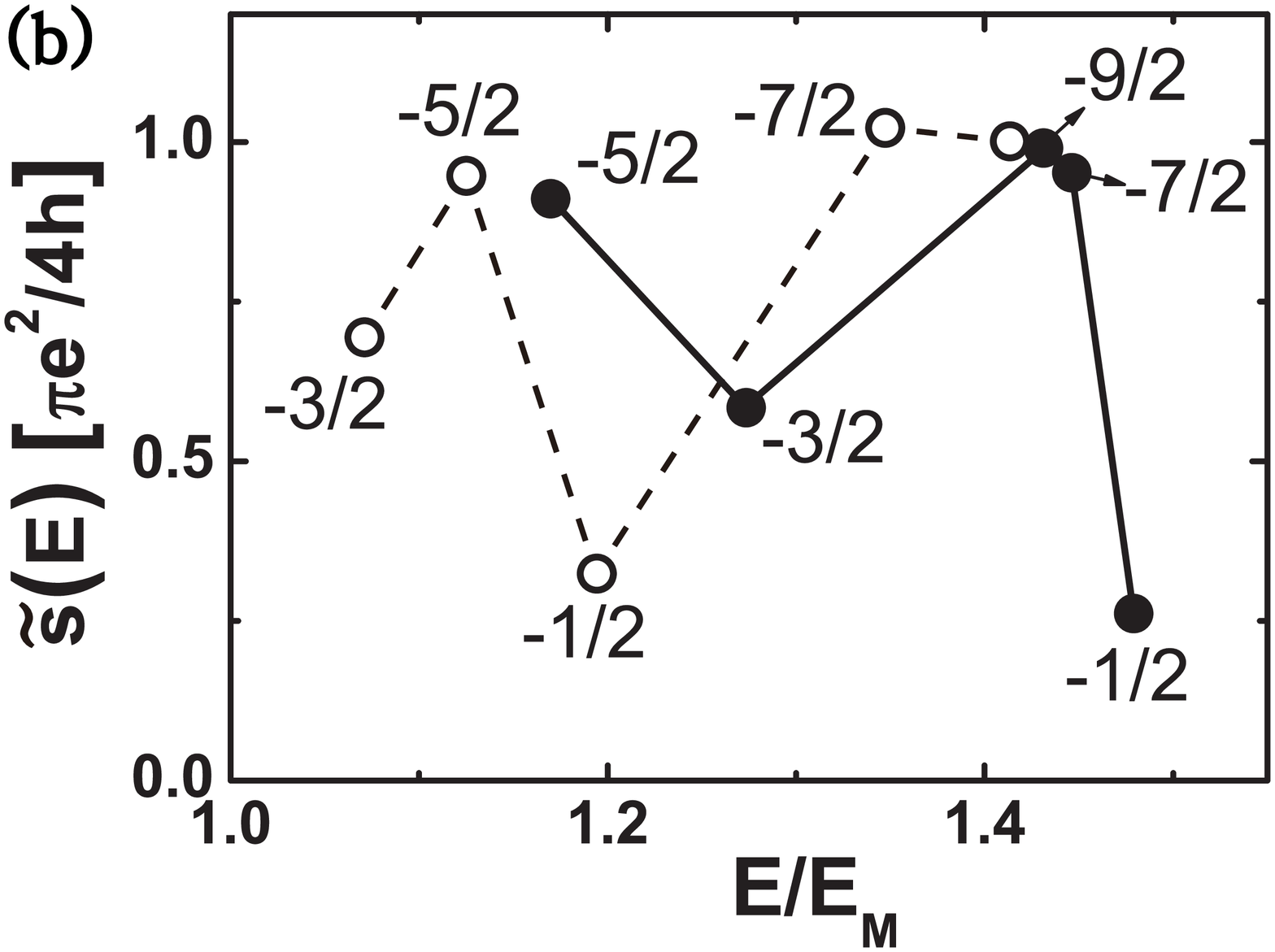}
\caption{ Scaled optical strength as a function of $E$ for
transitions between chiral and nonchiral states ($n=0\rightarrow
1$). Open circles represent optical strengths of single particle
transitions without many-body effects; the corresponding transitions
are indicated as arrows in Fig.\ref{anoma}(d). Optical strengths
renormalized by self-energy corrections are displayed by filled
circles.  Each circle is labeled by a fraction, which  denotes the
initial states $J$ of the optical transitions $(0,J)\rightarrow
(1,J+1)$. (a) $V_I=-7.9E_M$ and $R=0.3\ell$. (b) $V_I=-2E_M$ and
$R=\ell$.} \label{single}
\end{center}
\end{figure}

The scaled optical strength of Eq.(\ref{eq:scaled}) is given by
\begin{eqnarray}
\tilde{s}(E_{ex})=\frac{\pi}{2} \frac{e^2}{h} \frac{|\langle
1,J+1|\sigma_x|0,J\rangle|^2}{ \tilde{E}_{ex} },
\end{eqnarray}
where the scaled excitation energy is given by the difference
between renormalized impurity energies $\tilde{E}_{ex}=(E_{1,J+1}-
E_{0,J})/(\sqrt{2}E_M)$. The computed scaled optical strengths are
shown in Fig.\ref{single}. Let us first discuss the results in the
absence of self-energy corrections. When $(V_I/E_M,R/\ell)=(-2,1)$
the bare transition between anomalous states, $n=0\rightarrow 1$
with $J=-1/2\rightarrow 1/2$, has small  optical matrix element
$|\langle 0,-1/2|\sigma_{x}|1,1/2\rangle|^{2}=0.137$.   The
resulting value of the scaled optical strength is also small with
the value $0.325\frac{\pi e^2}{4h}$, see Fig.\ref{single}(b). For other bare transitions the optical matrix
elements are larger and their  scaled optical strengths are bigger
$\frac{1}{2} \frac{\pi e^2}{4h}<\tilde{s}(E_{ex})< \frac{\pi
e^2}{4h}$. Similar result also holds for $(V_I/E_M,R/\ell)=(-7.9,0.3)$, see
Fig.\ref{single}(a). We see that the values of the scaled optical strength are
almost unchanged by the exchange-self energy corrections (In
Fig.\ref{single} any two transitions labeled by same $J$ have
similar values of strengths). However, the corresponding
renormalized excitation energies display significant changes from
those of bare transitions.

\section{A filled Landau level}

\begin{figure}[!hbpt]
\begin{center}
\includegraphics[width=0.4\textwidth]{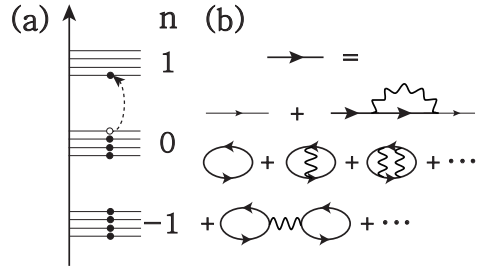}
\caption{ (a) Optical transition from a localized  state of the
filled $n=0$ LL  to a localized state of the empty $n=1$  LL. (b)
The self-energy is evaluated in a self-consistent HFA. Renormalized
excitation energy is computed within a time-dependent
self-consistent HFA including excitonic and depolarization effects.}
\label{fig:HF}
\end{center}
\end{figure}

\begin{figure}[!hbpt]
\begin{center}
\includegraphics[width=0.25\textwidth]{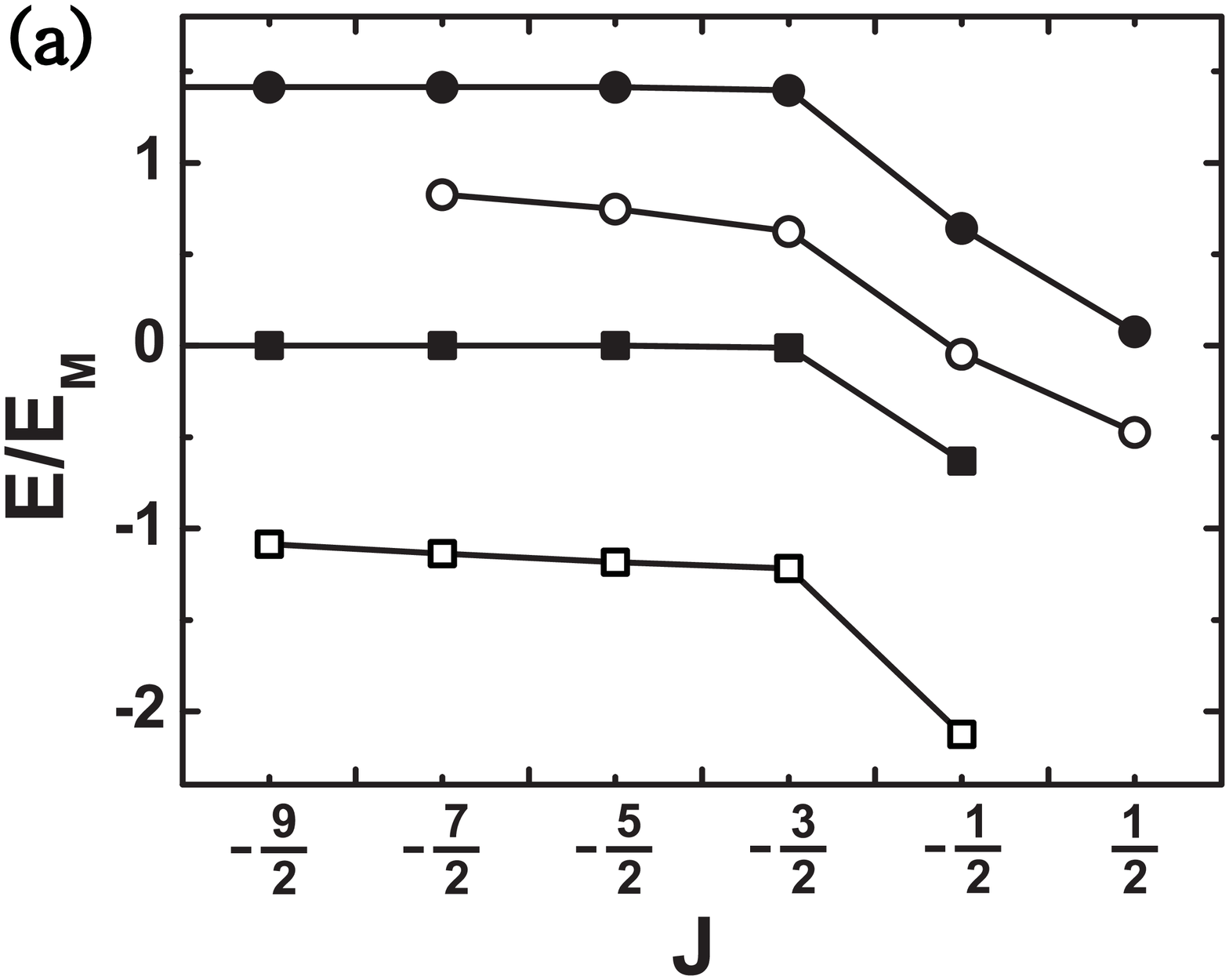}
\includegraphics[width=0.25\textwidth]{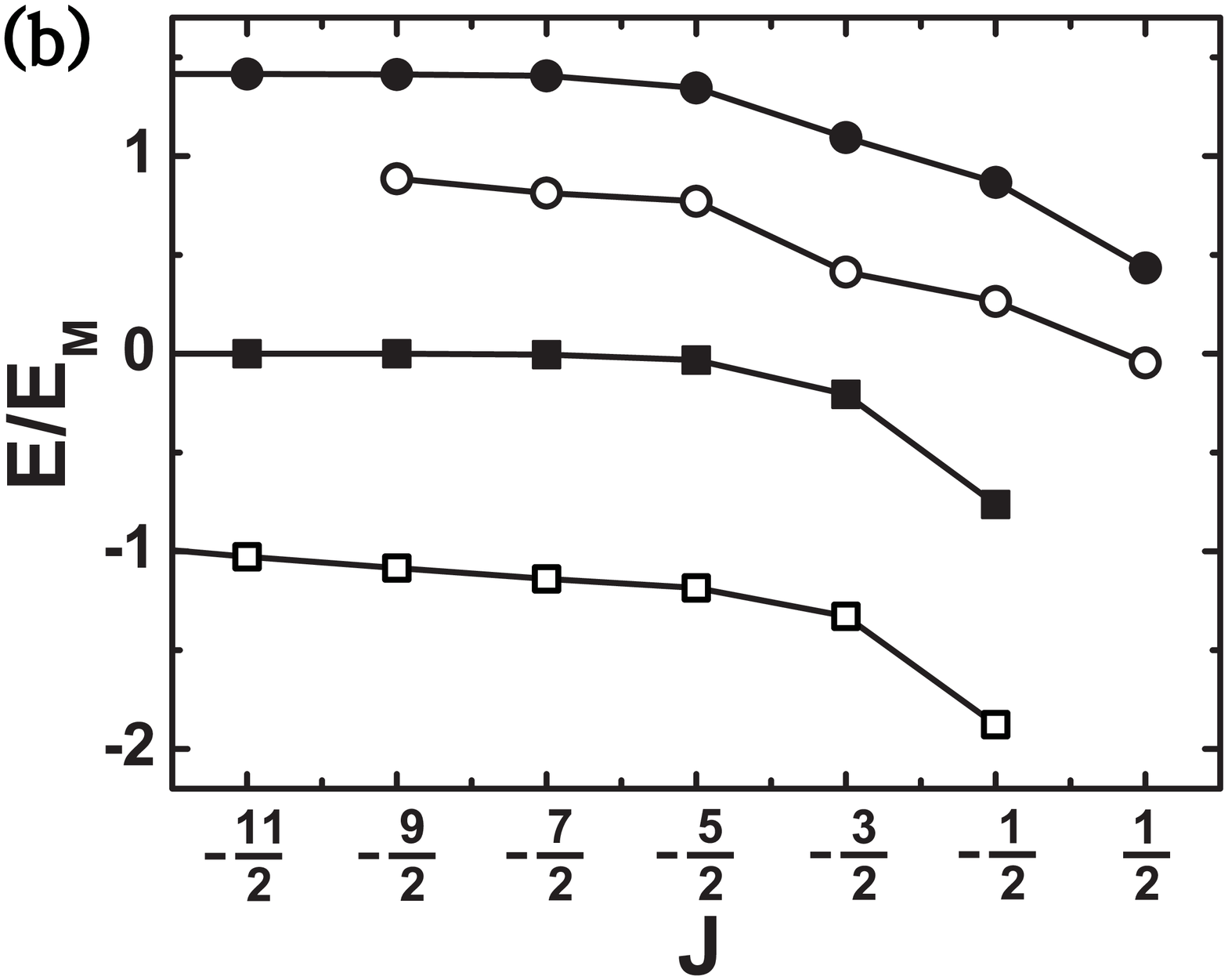}
\caption{ Bare impurity energies $\tilde{\epsilon}_{n,J}$  of $n=0$
(chiral) and $1$ (nonchiral) LLs by the impurity (filled symbols).
When the $n=0$ LL is completely filled these energies must be
corrected by including self-energies $E_{n,J}$ (open symbols). (a)
$V_I=-7.9E_M$ and $R=0.3\ell$. (b) $V_I=-2E_M$ and $R=\ell$.}
\label{fig:exself_imp}
\end{center}
\end{figure}

When a LL is completely filled both many-body correlations and self
energy effects may be  important.  Here we investigate how they may
affect the optical conductivity of anomalous states.  Here we assume
that $n=0,-1,-2,...$ LLs are filled. An optical transition leaves a
hole in the filled $n=0$ LL, see Fig.\ref{fig:HF}(a). Since there
can be several electron-hole excitations with $\Delta J=1$  an
eigenstate is given by a linear combination of these electron-hole
states. This implies that many-body correlation effects may be
important.  In this section we evaluate the magnitude of  the
excitonic and depolarization many-body effects (they are depicted in
Fig.\ref{fig:HF}(b)). Since $n=0$ LL is also filled in addition to
the $-1,-2,.....$  LLs the exchange self-energy acquires an
additional correction in comparison to the case of singly occupied
$n=0$ LL. The new exchange self-energies are shown in
Fig.\ref{fig:exself_imp}. In this section we explore whether these
features can affect the optical conductivity in a significant way.

\subsection{Many-body Hamiltonian matrix}

\begin{figure}[!hbpt]
\begin{center}
\includegraphics[width=0.3\textwidth]{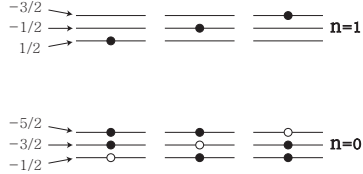}
\caption{ Examples of electron configurations that serve as basis
states of the Hamiltonian matrix. First, second, and third states
are labeled by $J=-1/2$, $-3/2$, and $-5/2$, respectively (see
Eq.(\ref{basis})).} \label{fig:basis}
\end{center}
\end{figure}
The electron-hole excitations form basis states of the Hamiltonian
matrix. Some of these configuration states are displayed in
Fig.\ref{fig:basis}. These electron configurations are coupled to
each other by many-body interactions. In a
TDHFA\cite{TDHF3,TDHF2,TDHF1} the Hamiltonian matrix elements can be
split into two parts
\begin{eqnarray}
H_{J,J'}=E_J^{d}\delta_{J,J'}+\Gamma_{J,J'}. \label{matrix}
\end{eqnarray}
The diagonal Hamiltonian matrix elements are
\begin{eqnarray}
E_J^{d}=E_{1,J+1}-E_{0,J}+\Gamma_{J,J}, \label{diagonal}
\end{eqnarray}
where the interaction energy of the  groundstate  $E_G$ is set to
zero. Note that it contains the contribution from the diagonal vertex corrections and that the renormalized single-particle energy $E_{n,J}$
contain self-energy corrections (now $n=0$  LL is also
filled, unlike the case studied  in Sec.3). The off-diagonal
Hamiltonian  elements  are the vertex corrections and are
 given by
\begin{eqnarray}
\Gamma_{J,J'}&=&-\langle 0,J;1,J+1|V|0,J';1,J'+1\rangle \nonumber\\
&& +\langle 0,J;1,J+1|V|1,J'+1;0,J'\rangle, \label{vertex}
\end{eqnarray}
where the first term is  the excitonic contribution
\begin{eqnarray}
-E_{ex}=-\langle 0,J;1,J+1|V|0,J';1,J'+1\rangle
\end{eqnarray}
and the second term is the  depolarization contribution
\begin{eqnarray}
E_{depol}=\langle 0,J;1,J+1|V|1,J'+1;0,J'\rangle .
\end{eqnarray}

\subsection{Excitation energies}

\begin{figure}[!hbpt]
\begin{center}
\includegraphics[width=0.25\textwidth]{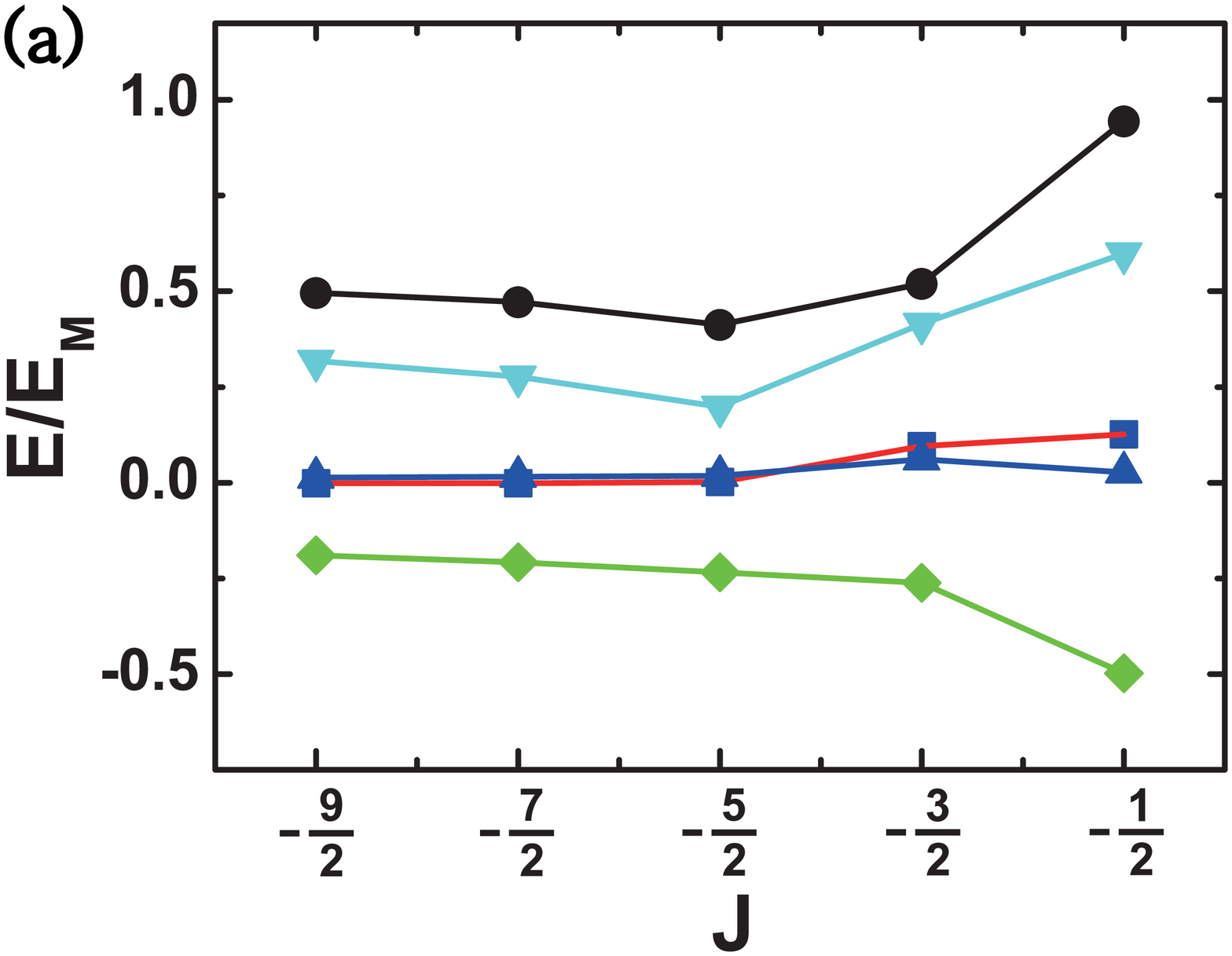}
\includegraphics[width=0.25\textwidth]{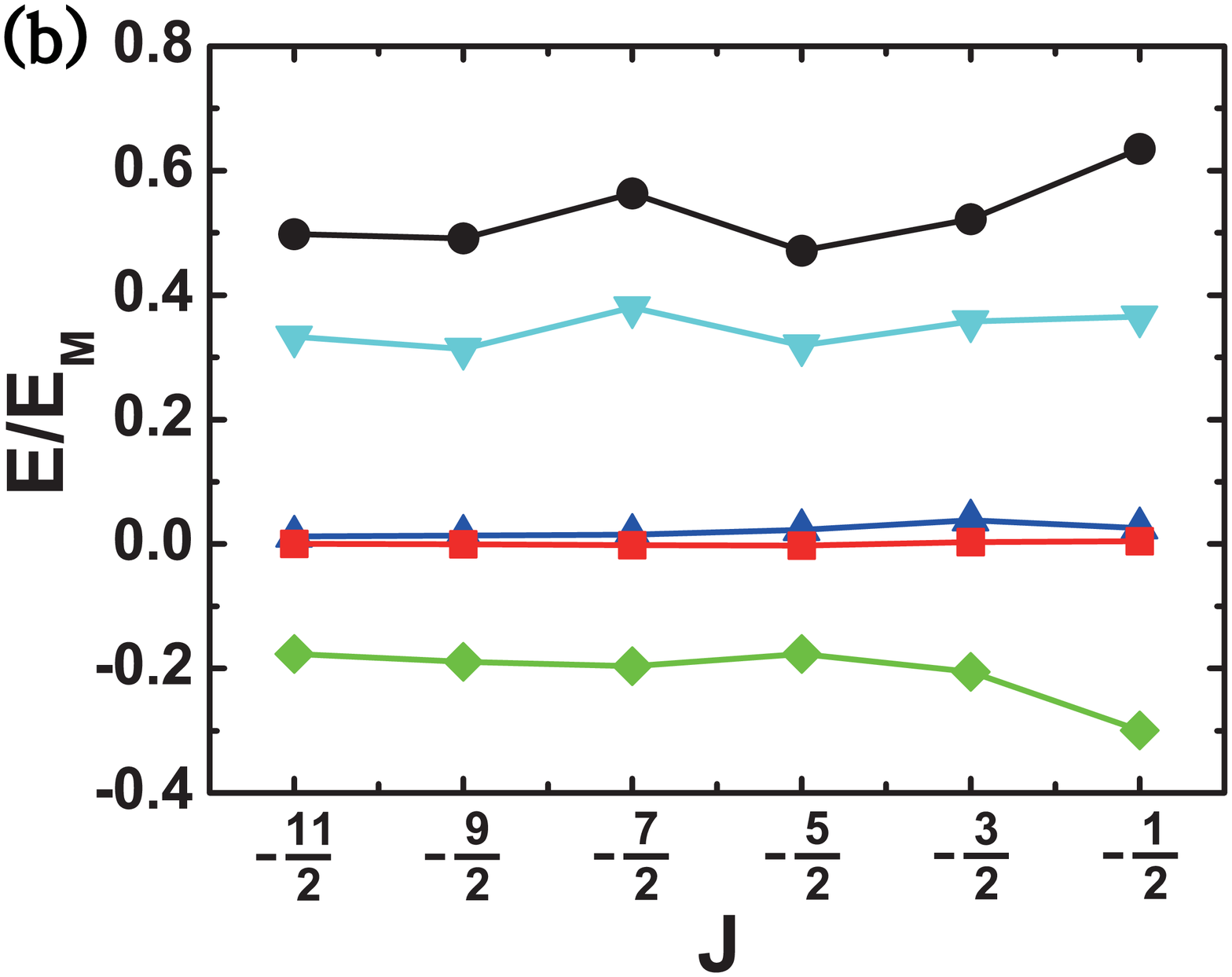}
\caption{  The following quantities in the diagonal elements of the
Hamiltonian matrix, see Eq.(\ref{diag}), are plotted as a function
of $J$: $\Sigma_{1,J}^X-\Sigma_{0,J}^X$ (circle),
$\Sigma_{1,J}^H-\Sigma_{0,J}^H$ (square),$-E_{ex}$ (diamond),
depolarization (triangle), and renormalization of the bare
excitation energy $\Sigma_{1,J}^X-\Sigma_{0,J}^X+\Gamma_{J,J}$,
ignoring the small Hartree corrections (inverted triangle). (a)
$V_I=-7.9E_M$ and $R=0.3\ell$. (b) $V_I=-2E_M$ and $R=\ell$.}
\label{fig:dia_imp}
\end{center}
\end{figure}

Despite the presence of  mixing between different electron
configurations, we find that the renormalized  {\it transition
energy} of $|0,J\rangle\rightarrow |1,J+1\rangle$ can be computed
approximately using a diagonal approximation: it is equal to
$E_J^{d}$, see Eq.(\ref{diagonal}), to
compute the values of optical strength accurately one may have to beyond the diagonal approximation. It can be broken
into various components
\begin{eqnarray}
H_{J,J}=E_{J}^{d}
&=&(\tilde{\epsilon}_{1,J+1}-\tilde{\epsilon}_{0,J})\nonumber\\
&+&(\Sigma_{1,J+1}^H-\Sigma_{0,J}^H)\nonumber\\
&+&(\Sigma_{1,J+1}^X-\Sigma_{0,J}^X)\nonumber\\
&+&(-E_{ex}+E_{depol}). \label{diag}
\end{eqnarray}
Values of these corrections  are plotted as a function of $J$ in
Fig.\ref{fig:dia_imp}. The exchange self-energy correction
$\Sigma_{1,J}^X-\Sigma_{0,J}^X$ is the most {\it dominant} term
(circles  in Fig.\ref{fig:dia_imp}). This diagonal approximation is physically equivalent to keeping only one electron-hole pair in the calculation.

\begin{figure}[!hbpt]
\begin{center}
\includegraphics[width=0.07\textwidth]{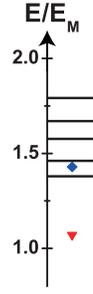}
\caption{Five eigenenergies obtained by diagonalizing $5\times5$
Hamiltonian matrix. Lowest bare impurity transition energy
$\tilde{\epsilon}_{1,1/2}-\tilde{\epsilon}_{0,-1/2}$ (inverted
triangle). Lowest transition energy in the diagonal approximation
(diamond) Parameters are $V_I=-2E_M$ and $R=\ell$.} \label{exact}
\end{center}
\end{figure}

Corrections to the transition energies beyond the diagonal
approximation can be investigated by solving the Hamiltonian matrix,
see Eq.(\ref{matrix}) (this method is equivalent to including the interaction between different electron-hole pairs). Fig.\ref{exact} displays obtained transition
energies using $5\times5$ Hamiltonian matrix for $V_I=-2E_M$. The
transition $(n,J)=(0,-3/2)\rightarrow (1,-1/2)$  as the lowest
renormalized transition energy  $1.38 E_{M}$ while  the
corresponding bare transition energy is $1.071E_M$. The deviation
between the two values is $22.4\%$. Note that the many-body effects
{\it increase} the transition energy from the bare value. For this
transition the corresponding many-body state has the expansion
coefficients $C_{J}=(-0.272,-0.733,-0.594,-0.161,-0.097)$,  given by
the linear combination Eq.(\ref{linear-comb}).  There is a
substantial mixing between different electron-pair configurations.
Note that $C_J$ for $J=-3/2$ and $-5/2$ are dominant.  In the
diagonal approximation the transition energy is $1.46E_M$, which is
close to the renormalized value.  So despite strong mixing the
diagonal approximation given a good estimate of transition energies.

\subsection {Results of optical conductivity}

\begin{figure}[!hbpt]
\begin{center}
\includegraphics[width=0.25\textwidth]{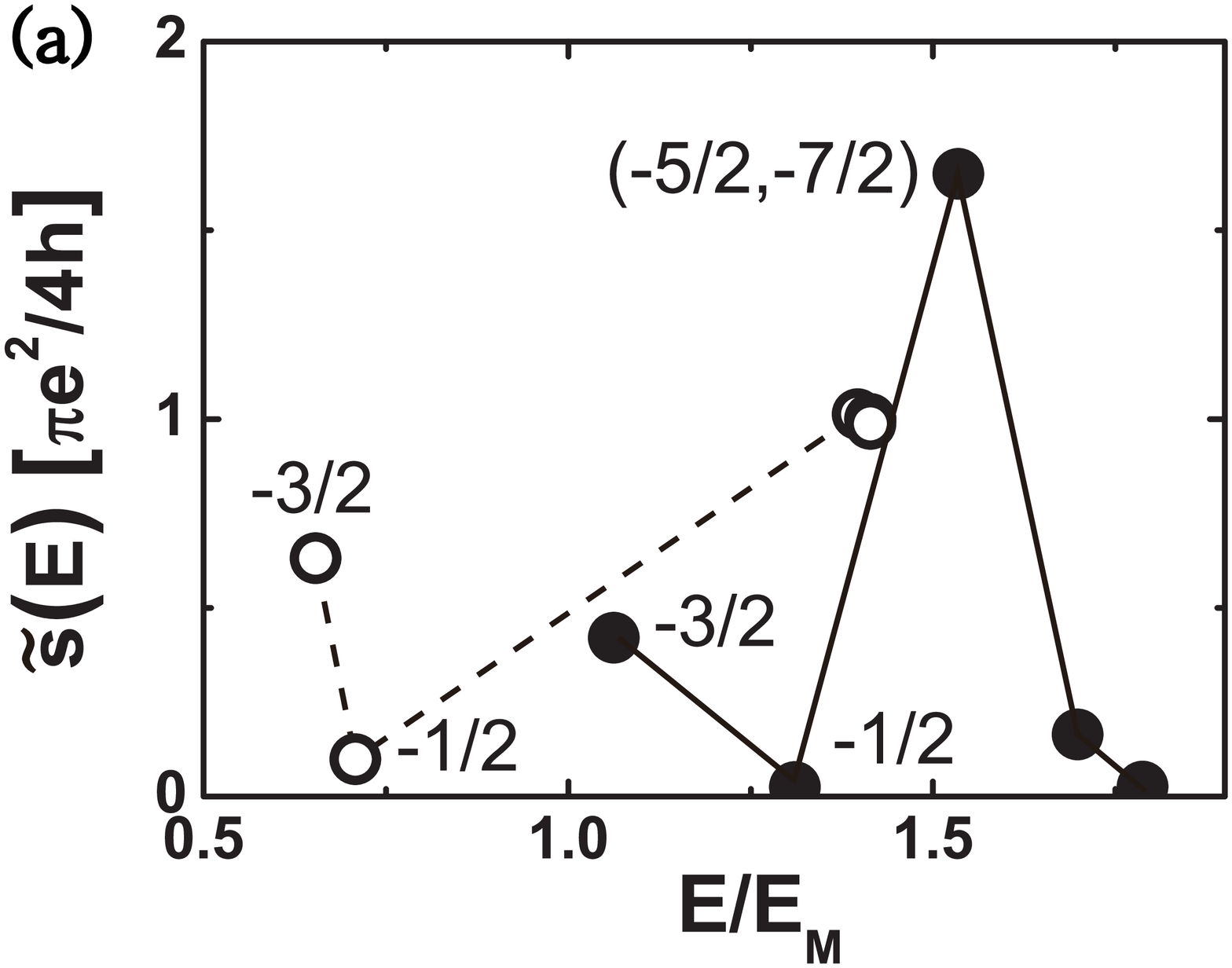}
\includegraphics[width=0.25\textwidth]{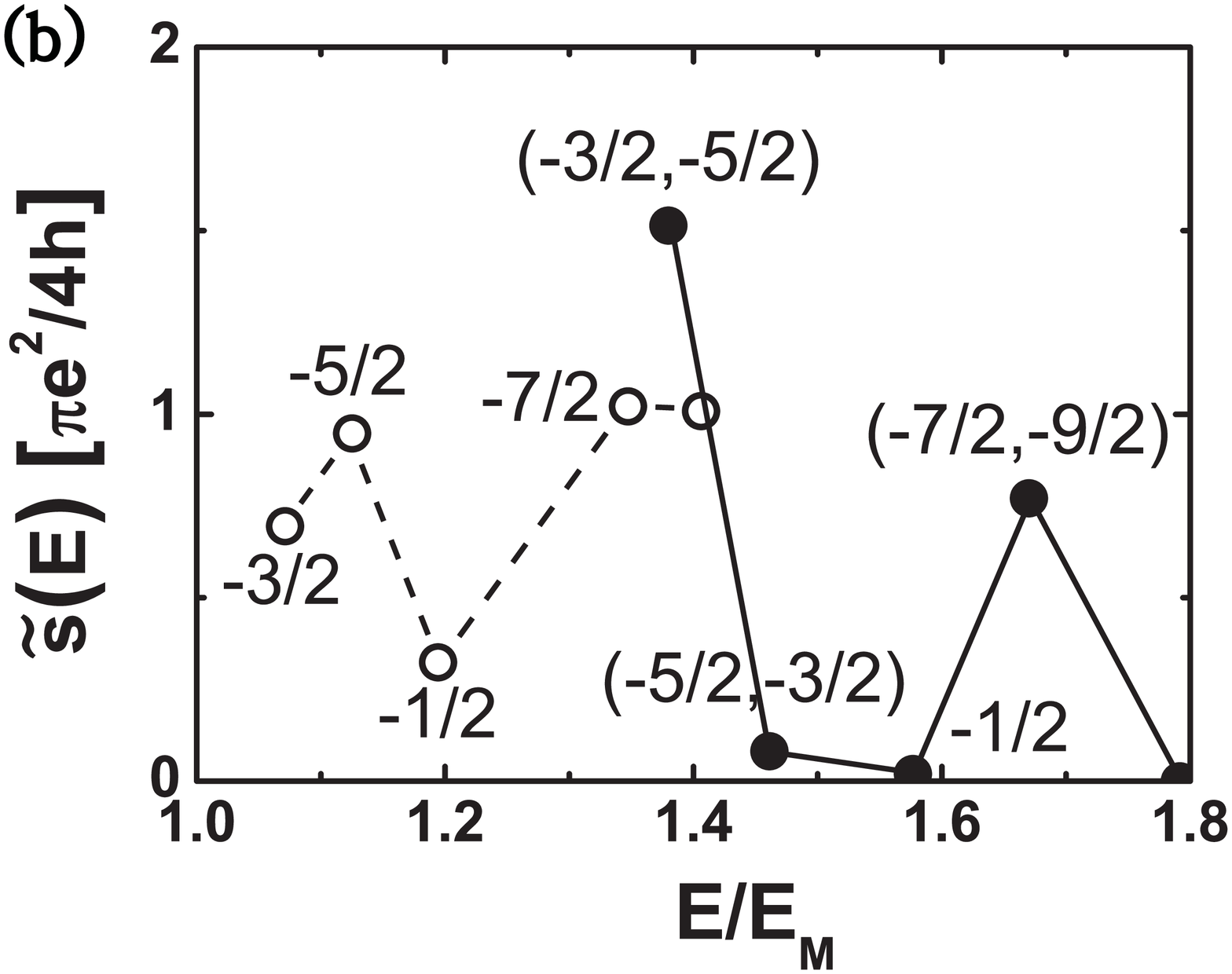}
\caption{ Scaled optical strength as a function of $E$ for
transitions between chiral and nonchiral states ($n=0\rightarrow
1$). Open circles represent optical strengths of single particle
transitions without many-body effects; the corresponding transitions
are indicated as arrows in Fig.\ref{anoma}(d).  In this figure  a
set of numbers $(-3/2,-5/2)$ means that the many-body excited state
is dominantly a mixture electron-hole pair states
$|\psi_{-3/2}\rangle$ and $|\psi_{-5/2}\rangle$ (electron-hole pair
states $|\psi_{J}\rangle$ are defined in Eq.(\ref{basis})). On the
other hand, a transition labeled by a single value of $J$ means that
the electron-hole pair state $|\psi_{J}\rangle$ is the most dominant
one in the linear combination given by  Eq.(\ref{linear-comb}).
Optical strengths renormalized by self-energy and vertex corrections
are displayed by filled circles. (a) $V_I=-7.9E_M$ and $R=0.3\ell$.
(b) $V_I=-2E_M$ and $R=\ell$.} \label{optic}
\end{center}
\end{figure}

Let us show the result of the optical conductivity obtained by
including  full many-body effects.  We diagonalize $5\times5$
Hamiltonian matrices (here we are mostly interested in lower energy
excitations than  that of magnetoplasmons; a significantly larger
Hamiltonian matrix is needed to describe magnetoplasmon physics).
The many-body eigenstates are given by the linear combination of
electron-hole pair states $ |\psi_{J}\rangle$ (see
Eq.(\ref{linear-comb})). The obtained result for the scaled optical
strength, given by Eq.(\ref{eq:scaled}), is displayed in
Fig.\ref{optic} together with the bare values in the absence of
many-body effects.

A transition between two  anomalous impurity states is labeled by
$-1/2$, see Figs.\ref{optic} (a) and (b). It is created by the
transition between two impurity anomalous states $|0,-1/2\rangle$
and $ |1,1/2\rangle$, see Fig.\ref{anoma}. Our numerical work shows
that it involves dominantly only one electron-hole pair state
$|\psi_{-1/2}\rangle$, and is nearly uncorrelated,  displaying a
{\it small mixing} with other electron-hole pair states. We
investigate the physical origin of this effect  by treating the
off-diagonal matrix elements of the many-body Hamiltonian,
Eq.(\ref{vertex}), as a perturbation and computing the first-order
correction to the many-body state. The degree of mixing between the
electron-hole pair states $|\psi_{-1/2}\rangle$ and
$|\psi_{-3/2}\rangle$ is small since the magnitude of vertex
correction between them $\Gamma_{-1/2,-3/2}$ is smaller than the
{\it difference} between their renormalized transition energies
$|E_{-3/2}^{d}-E_{-1/2}^{d}|$ ($E_{J}^{d}$ is computed within the one electron-hole pair approximation and is given in
Eq.(\ref{diagonal})).  Although  the optical conductivity of this
transition  is small,  it  can be {\it enhanced} by tuning the value
of magnetic field (see Table 1).

The other transitions between a normal and an  anomalous state are
labeled by $-3/2$ and $(-3/2,-5/2)$, see Figs.\ref{optic} (a) and
(b). These transitions are created by the optical transition between
impurity states $|0,-3/2\rangle \rightarrow |1,-1/2\rangle $ (note
that only $|1,-1/2\rangle $ is an anomalous state, see
Fig.\ref{anoma}). It is nearly uncorrelated for the parameter values
$(V_I,R/\ell)=(-7.9,0.3)$ (see Fig.\ref{optic}(a)). However,  this
same state is more correlated for $(V_I,R/\ell)=(-2,1)$, resulting
in the  state $(-3/2,-5/2)$, see Fig.\ref{optic}(b). This correlated
state has a significantly enhanced value of the optical strength in
contrast to the state $-3/2$.   The interplay between the vertex
corrections and the renormalized transition energies can thus affect
substantially  the optical transitions involving one  anomalous
state.

\section{Summary and discussion}

We have investigated a new feature in the impurity cyclotron
resonance  that is common to various localized potentials of
graphene. Application of a magnetic field makes $n=0$ and $n=1$ LL
states chiral and nonchiral, respectively.   This  has a non-trivial
implication, leading to only one anomalous state for the $n=0$ LL
while two for the $n=1$ LL, see Fig.2.  It  is a unique feature of
physics of finite magnetic fields.  The anomalous boundstates  are
strongly localized inside the well with a broad peak outside the
well with width comparable to the magnetic length. In this paper we
have proposed that anomalous boundstates may exhibit an unusually
small value of magneto-optical conductivity since  optical matrix
elements of anomalous states are negligible compared to those of
ordinary states. The effect of many-body interactions on their
optical conductivity is investigated for a completely filled LL
using a self-consistent TDHFA. We find that an excited electron-hole
state originating from the optical transition between two anomalous
impurity states exhibits small correlations with other electron-hole
states, despite it displaying substantial exchange self-energy
effects. This is a consequence of a small vertex correction in
comparison to the difference between renormalized transition
energies computed within the one electron-hole pair approximation.  We  find that many-body interactions do not enhance the
strength of its  optical conductivity. However, by tuning the value
of magnetic field its strength may be enhanced significantly. There
is also a transition between a normal and an anomalous  impurity
states. Unlike the optical transition between two anomalous state,
we find, in this case, that the optically created electron-hole pair
can be substantially correlated with other electron-hole pairs, and
that its optical strength can be significant.

Note that the eigenenergies of a parabolic potential  are complex,
implying that the lifetime in the potential is finite. For optical
studies states with small  imaginary part of the eigenenergies are
desirable.  If the confining  potentials vary fast there may be some
valley mixing, which leads to splitting of eigenenergies\cite{Park2}.  A
tight-binding calculation can be used to investigate this effect.

Recently several  infrared absorption experiments of graphene have
been performed\cite{Gus,Fal,Li}. We suggest that this type of
experiment  be performed in magnetic fields on donor impurities or
on quantum dot arrays in graphene, just like the case of
two-dimensional massful electrons\cite{Jar,Gold,Heit}. It would be
interesting to observe anomalous transitions in the impurity
cyclotron resonance  in the regime $R/\ell<1$, and confirm the
sensitive dependence of their optical strength on magnetic field. In
this paper we considered donor impurities. For acceptors or antidots\cite{Park2}
we can use the transformation $V(r)\rightarrow -V(r)$ with the
eigenenergies $E\rightarrow -E$ (eigenstates are unchanged).

\ack This research was supported by Basic Science Research Program
through the National Research Foundation of Korea(NRF) funded by the
Ministry of Science, ICT $\&$ Future Planning(MSIP) (NRF-2012R1A1A2001554).
In addition this research was supported by a Korea University Grant.


\section*{References}
\begin{thebibliography}{10}
\bibitem{Mac0} A. K. Geim and A. H. MacDonald, Phys. Today
{\bf 60}, 35 (2007).

\bibitem{Ben} C. W. J. Beenakker, Rev. Mod. Phys.  {\bf 80}, 1337 (2008).

\bibitem{Castro} A. H. Castro Neto, F. Guinea, N. M. R. Peres, K. S. Novoselov, and A. K. Geim, Rev. Mod. Phys. {\bf 81}, 109 (2009).

\bibitem{Jac} R. Jackiw, 'Delta function potentials in two- and three-dimensional
quantum mechanics', in M.A.B. B\'{e}g Memorial Volume, eds. A. Ali and P.
Hoodbhoy (World Scientific, Singapore, 1991).


\bibitem{com0} For massful
electrons  kinetic term scales as $1/r^2$. The potential is
comparable to or weaker  than the kinetic term.

\bibitem{Mat} A. Matulis and  F. M. Peeters, Phys. Rev. B  {\bf 77}, 115423 (2008);
P. G. Silvestrov and K. B. Efetov, Phys. Rev. Lett.  {\bf 98},
016802(2007).



\bibitem{gia}G. Giavaras, P. A. Maksym, and M. Roy, J. Phys.: Condens. Matter {\bf
21}, 102201 (2009).
\bibitem{Rec} P. Recher, J. Nilsson, G. Burkard, and B. Trauzettel, Phys. Rev. B
{\bf 79}, 085407 (2009).

\bibitem{Sch}S. Schnez, K. Ensslin, M. Sigrist, and T. Ihn, Phys. Rev. B
{\bf 78}, 195427 (2008).


\bibitem{Park} P. S. Park, S. C. Kim, and S. -R. Eric Yang, Phys. Rev. B  {\bf 84}, 085405 (2011).


\bibitem{Kim0} S. C. Kim and S. -R. Eric Yang,   Annals of
Physics {\bf 347}, 21 (2014).  In the case of the impurity  Coulomb
potential a  regularization parameter $R$ must be introduced:
$V(r)=-\frac{e^2}{\epsilon R}$ for $r<R$ and $-\frac{e^2}{\epsilon
r}$ for $r>R$ ($\epsilon$ is the dielectric constant). When the
strength of the potential is strong this is needed to prevent the
spurious effect of Coulomb fall to the center of the potential.
\bibitem{Ho} C. L. Ho and V. R. Khalilov, Phys. Rev. A  {\bf 61}, 032104
(2000);
 Y. Zhang, Y. Barlas, and K. Yang, Phys. Rev. B  {\bf 85}, 165423
(2012).

\bibitem{Kim2} The natural length scale of a parabolic potential $\frac{1}{2}\kappa r^2$ is $R=(\frac{\hbar v_F}{\kappa})^{1/3}$.
S. C. Kim, J.  W. Lee,  and S. -R. Eric Yang, J. Phys.: Condens.
Matter {\bf 24}, 495302  (2012);  P. S. Park, S. C. Kim, and S. -R.
Eric Yang, Phys. Rev. Lett.  {\bf 108}, 169701 (2012).





\bibitem{Jar} N. C. Jarosik, B. D. McCombe, B. V. Shanabrook,  J. Comas, J.
Ralston, and G. Wicks,  Phys. Rev. Lett. {\bf 54}, 1283 (1985).

\bibitem{Gold} V. J. Goldman, H. D. Drew, M. Shayegan, and D. A. Nelson, Phys. Rev. Lett. {\bf 56}, 968
(1986).

\bibitem{zero}
T. Yamamoto, T. Noguchi, and K. Watanabe, Phys. Rev. B  {\bf
74}, 121409(R) (2006); M. Zarenia, A. Chaves, G. A. Farias, and F.
M. Peeters, Phys. Rev. B  {\bf 84}, 245403 (2011).



\bibitem{many} Optical properties of  bulk graphene at $B=0$ have been investigated using
various many-body techniques: D. Prezzi, D. Varsano, A. Ruini, A.
Marini, and E. Molinari, Phys. Rev. B {\bf 77}, 041404(R) (2008); L.
Yang, J. Deslippe, C. -H. Park, M. L. Cohen, and S. G. Louie, Phys.
Rev. Lett. {\bf 103}, 186802 (2009);  W. Wei and T. Jacob, Phys.
Rev. B {\bf 87}, 115431 (2013).


\bibitem{com1}  The dimensionless strength of the electron-electron Coulomb interaction is given by
the ratio between the energy scales for the Coulomb interaction and
LL energy separation $ g=(\frac{e^2}{\epsilon \ell})/(\frac{\hbar
v_F}{\ell})=\frac{e^2}{\epsilon\hbar v_F} $, and note that this
dimensionless coupling constant is independent of magnetic field.
Hereafter we will set $g=0.5$.

\bibitem{TDHF3} Y. A. Bychkov,
S. V. Iordanskii, and G. M. \'{E}liashberg, Pis'ma Zh. Ekps. Teor. Fiz. {\bf
33}, 152 (1981) [JETP Lett. {\bf 33}, 143 (1981)].

\bibitem{TDHF2}  C. Kallin and
B. I. Halperin, Phys. Rev. B {\bf 30}, 5655 (1984).

\bibitem{TDHF1} A. H. MacDonald, J. Phys. C {\bf 18}, 1003 (1985).



\bibitem{Bychkov} Y. A. Bychkov and G.
Martinez, Phys. Rev. B {\bf 77} 125417 (2008); R. Rold\'{a}n, J. -N. Fuchs, and M. O. Goerbig, Phys. Rev. B {\bf 82} 205418 (2010).


\bibitem{Pfa} D. Pfannkuche, V. Gudmundsson, and P. A. Maksym, Phys. Rev. B {\bf
47}, 2244 (1993).  In this paper, just like in our approach, the
states of a quantum-dot in a magnetic field are obtained exactly
before applying Hartree-Fock method.


\bibitem{Yosi}  D. Yoshioka, The Quantum Hall Effect (Springer, Berlin, 1998).
In the absence of a potential these functions are related to
Laguerre polynomials.



\bibitem{optic} In the presence of a strong impurity potential the selection rule
$\Delta n=1$  must be relaxed since different LLs get mixed.
However, the corresponding transitions have small  matrix elements
and will be ignored in the following.

\bibitem{Yang} S. -R. Eric Yang and A. H. MacDonald, Phys. Rev. B {\bf 42}, 10811
(1990).

\bibitem{Lee} J. W. Lee, S. C. Kim, and S. -R. Eric Yang, Solid State Commun., {\bf 152}, 1929 (2012).
In this paper exchange self-energy is computed when $\mathbf{K}$ and
$\mathbf{K'}$ valleys are coupled.

\bibitem{Park2} P. S. Park, S. C. Kim, and S. -R. Eric Yang, J. Phys.: Condens. Matter {\bf 22}, 375302 (2010).


\bibitem{Gus} V. P. Gusynin, S. G. Sharapov, and J. P. Carbotte, Phys. Rev. Lett. {\bf 96}, 256802 (2006);
ibid. {\bf 98}, 157402 (2007); J. Phys.: Condens. Matter {\bf 19},
026222 (2007).

\bibitem{Fal} L. A. Falkovsky and A. A. Varlamov, Eur. Phys. J. B {\bf 56}, 281 (2007);
L. A. Falkovsky and S. S. Pershoguba, Phys. Rev. B {\bf 76}, 153410
(2007).

\bibitem{Li} Z. Q. Li, E. A. Henriksen, Z. Jiang, Z. Hao, M. C. Martin, P. Kim, H. L. Stormer, and D. N. Basov, Nature Phys. {\bf 4}, 532 (2008);
 K. F. Mak, M. Y. Sfeir, Y. Wu, C. H. Lui, J. A. Misewich, and T. F. Heinz, Phys. Rev. Lett. {\bf 101}, 196405 (2008).

\bibitem{Heit} D. Heitmann and J. P. Kotthaus, Phys. Today {\bf 46}, 56
(1993).











\end{thebibliography}
\end{document}